\documentclass[]{aastex631}

\usepackage{amsmath,comment}

\begin{document}

\title{Experimental Validation for Serial Conjunction of Diffraction-limited Coronagraph and Fiber Nuller}

\author[0000-0003-2690-7092]{Satoshi Itoh}
\author{Taro Matsuo}
\author{Reiki Kojima}
\affiliation{Department of Particle and Astrophysics\\ Graduate School of Science\\ Nagoya University, Furocho, Chikusa-ku, Nagoya, Aichi, 466-8602, Japan}
\author{Takahiro Sumi}
\affiliation{Faculty of Earth and Space Science\\ Osaka University 1-1, Machikaneyama-cho, Toyonaka, Osaka 560-0043, Japan}
\author{Motohide Tamura}
\affiliation{Department of Astronomy, Graduate School of Science, The University of Tokyo, 7-3-1 Hongo, Bunkyo-ku, Tokyo 113-0033}
\affiliation{Astrobiology Center, 2-21-1 Osawa, Mitaka, Tokyo 181-8588, Japan}
\affiliation{National Astronomical Observatory of Japan, 2-21-1 Osawa, Mitaka, Tokyo 181-8588, Japan} 



\begin{abstract}
We report the experimental results of a serially conjoined nuller system, which combines a type of Lyot coronagraph with a fiber nuller. The utilized one-dimensional diffraction-limited coronagraph (1DDLC) has promising features (binary nuller, small inner working angles (IWAs)). Still, it has a performance that is highly sensitive to spectral bandwidth and tilt aberrations. Nevertheless, for the 1DDLC, wavelengths other than the design wavelength introduce leaks with a flat wavefront on the Lyot-stop plane, preserving the same complex amplitude profile as an on-axis point source. This property supports the concept of serially coupling additional nullers after the 1DDLC. The fiber-nulling unit employs a Lyot-plane mask, relay optics (1/100×), and a single-mode fiber. The Lyot-plane mask splits the incoming beam—comprising leakage from the 1DDLC and planetary light—into four beams so that, in principle, the on-axis single-mode fiber does not couple with the on-axis leak from the 1DDLC. For the wavelength 6-\% less than the coronagraph's design-center wavelength, we confirmed the contrast mitigation ability of $3.5 \times 10^{-5}$, which is about 1/20 times the value of the case with only 1DDLC. The resultant value approximately reaches the 1DDLC's contrast mitigation ability at the design center wavelength demonstrated in the previous study, suggesting that the combined system works robustly against the broad spectral bandwidth. Future work needs to address the demonstration of the anticipated robustness for the contrast-mitigation level lower than about $10^{-5}$.
\end{abstract}



\section{Introduction} \label{sec:intro}
We use coronagraph to observe dim objects at narrow angular separations from bright sources, such as solar corona \citep[e.g.,][]{1933JRASC..27..225L,2024JKAS...57..183P}, circumstellar structures \citep[e.g.,][]{1977Icar...30..422K,1994ApJ...428..797N,1997MNRAS.292..896M,2024ApJ...977..247L,2024Natur.633..789M,2025AJ....169...17B}, and structures around quasars \citep[][]{2024A&A...683L...5R}.
Many of the coronagraphs belong to the Lyot coronagraph \citep[e.g.,][]{1997PASP..109..815R,2002ApJ...570..900K,2005ApJ...633.1191M,2005OptL...30.3308F}, which consists of two main elements: focal-plane mask and pupil-plane aperture referred to as Lyot stop to separate the targets and the contamination sources before light detection, sometimes cooperating with precise wavefront control.
The photonic integral circuit (PIC)-based coronagraph \citep[][]{2024SPIE13092E..1TS} serves similar objectives as the Lyot coronagraphs with revolutionary compact devices in principle but has maturation levels of the current technology lower than those of the Lyot coronagraphs. 
The fiber nulling methods \citep[e.g.,][]{2022SPIE12180E..0NS,2024OExpr..3219924S,2024ApJ...965L..15E} use the fiber-coupling characteristic to separate the stellar (contamination) and planetary (target) light with a diffraction-limited separation angle.
Future applications of these coronagraph technologies include the ``search for life'' on exoplanets \citep[e.g.,][]{DesMarais+2002,Seager+2016,Kaltenegger+2017,Fujii+2018,2024JATIS..10c5004M,2024JATIS..10c4006S,2024A&A...683L...5R}.

The one-dimensional diffraction-limited coronagraph \citep[1DDLC,][]{Itoh+2020,2023PASP..135f4502I} recently joined the Lyot coronagraphs. 
The 1DDLC uses a focal-plane mask with mask function values from minus one to one along the real axis on the complex number plane. 
Since $-1=e^{\pi i}$, the minus sign in the mask function means the $\pi$-rad phase modulation. 
Hence, this mask function includes a $\pi$-rad phase modulation and amplitude modulation. 

A mathematical theory \citep[][]{Itoh+2020} similar to that of the band-limited mask coronagraph \citep[][]{2002ApJ...570..900K} derives the 1DDLC focal-plane mask pattern.
From the fact that the 1DDLC focal-plane mask has negative values ($\pi$-rad modulation) on the region with a width corresponding to about a half-width of the on-axis (stellar) point-spread function (PSF), we can say that the 1DDLC's operational principle resembles that of the complex mask Lyot coronagraphs \citep[][]{Guyon+2010}.
The 1DDLC's focal-plane mask reshapes only the focal amplitude of the on-axis source to a high-spatial-frequency spatial wavelet diffracted toward outside the Lyot-stop.
\citet[][]{2023PASP..135f4502I} demonstrates an example of how to implement the 1DDLC focal-plane mask, which resembles the implementation of the vector vortex mask coronagraph \citep[][]{2010ApJ...709...53M}.

The theory on the 1DDLC anticipates the following promising facts: 
(i) the 1DDLC has a one-dimensional focal-plane mask pattern, therefore it serves as a binary-star nuller, and (ii) the 1DDLC can perfectly erase on-axis monochromatic point sources in the sky while it transmits off-axis sources, which have the diffraction-limited separation angles with respect to the on-axis sources, from the sky to the focal plane.
The 1DDLC's performance varies sensitively to the wavelength deviation from the design center wavelength of light and the tilt aberration (i.e., telescope pointing jitters or stellar angular diameters).
Nevertheless, the effect of the wavelength deviation brings only a ``clean''  stellar leak.
Here, the ``clean'' leak means the stellar leak with a constant amplitude and phase on the Lyot stop aperture.
From the definition, the ``clean'' leak has the same functional profile as the amplitude of the on-axis point source before the coronagraph focal-plane mask except for a constant multiplication.
This nature encourages the ``serial conjunction'' of other nullers after the 1DDLC.

In this study, we report the results of a first preliminary experiment of a serially conjoined coronagraph system \citep[][]{2024AJ....167..235I} composed of the 1DDLC and a fiber nulling.
This experiment aims to demonstrate that the serial connection yields an expansion of the spectral bandwidth compared to the case with only the 1DDLC.
\begin{figure*} 
    \centering
    \includegraphics[width=0.9\linewidth]{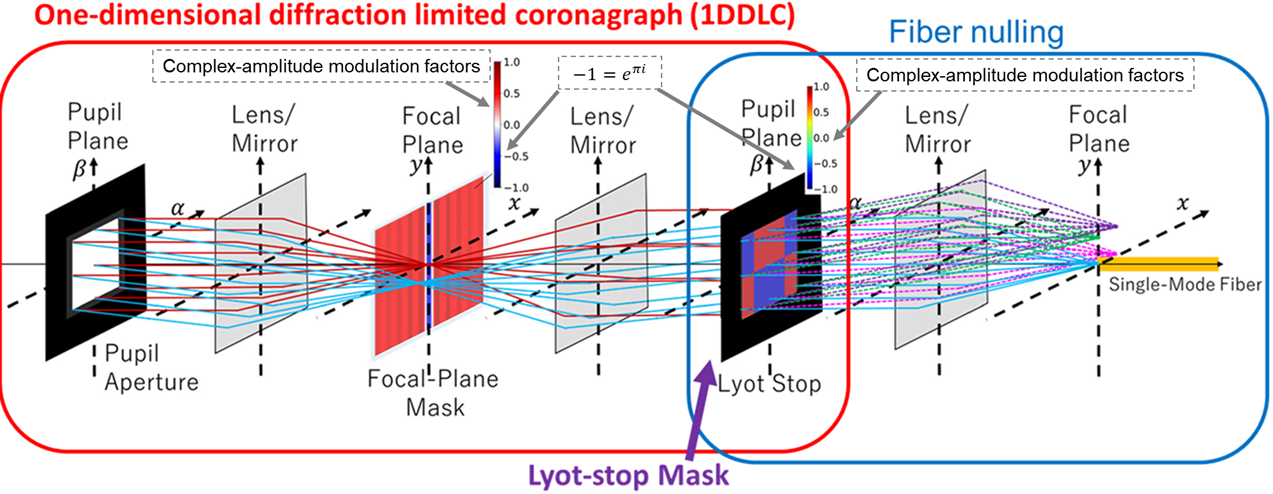}
    \caption{Schematic for the coronagraph system that consists of serial conjunction of the 1DDLC and the fiber nulling.}
    \label{fig:SC}
\end{figure*}
Figure $\ref{fig:SC}$ illustrates the concept of the system, where light goes through the coronagraph unit and then the fiber-nulling unit.
At the Lyot plane, we locate a Lyot-plane achromatic phase mask that takes the mask-function values of one ($=e^{0 i}$)  and minus one ($=e^{\pi i}$) using the geometrical phase of light.
Single-mode fibers located at the origin of the coordinate have no transmittance for the odd-function components of amplitude-spread functions of light. 
To utilize this nature of single-mode fibers, the Lyot-plane mask changes the 1DDLC's stellar leak (due to two different causes: wavelength deviation from the mask's design-center wavelength and the least-order tilt aberration) to an odd function for at least one variable among two independent variables of the Cartesian coordinates on the Lyot-stop plane. 
This method works because the Fourier transform that expresses the propagation of the light wave between the pupil and focal planes preserves the parity (odd/even) of the amplitude spread functions.  
Function patterns with a parity desired for mitigating the stellar light include uncountable variation.
Among them,  \citet[][]{2024AJ....167..235I} chose the rectangular-wave pattern with a spatial period same as the pupil width to secure the throughput of the planet with the diffraction-limited separation angles as much as possible. 
For an ideal optical fiber, the off-axis sources have a theoretical coupling efficiency of 22\% at diffraction-limited separation angles. 
In Section \ref{sec:meth}, we explain the methods of the experiment including the experimental setup (Section \ref{sec:meth1}) and the data acquisition procedure (Section \ref{sec:meth2}).
Section \ref{sec:resu} exhibits the experiment result.
Section \ref{sec:conc} summarises the present study.


\section{Method} \label{sec:meth}
Here, we explain how we conduct a first preliminary experiment of the serially conjunct coronagraph system of the 1DDLC and the fiber nulling.

\subsection{Experimental Setup} \label{sec:meth1}
\begin{figure}[htb] 
    \centering
    \includegraphics[width=1.0\linewidth]{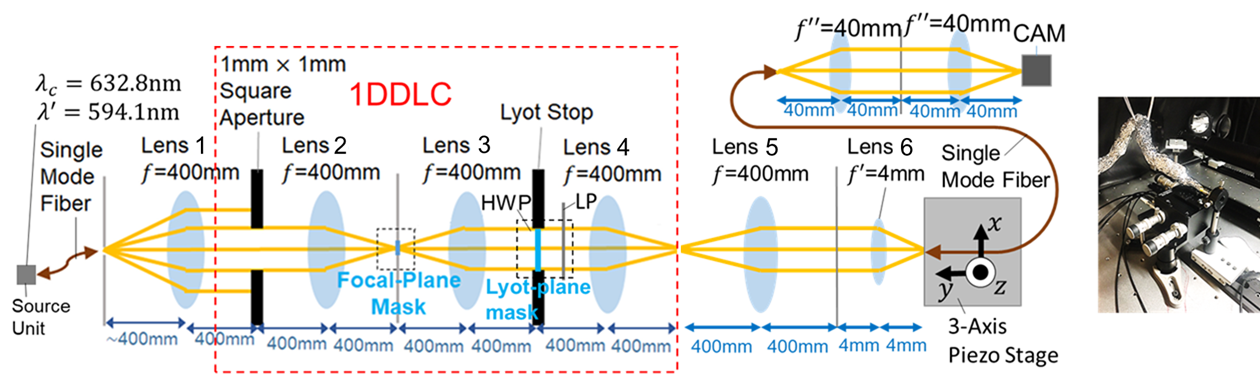}
    \caption{Illustrated optical setup of the experimental setup. The orange lines indicate the approximated light paths assuming no focal-plane masks and no Lyot-plane masks. We write the axes of 3-dimensional Cartesian coordinates on the 3-axis piezo stage to define the driving axes of the stage. (right) The picture of the lens 6, the 3-axis piezo stage, and the single-mode fiber. }
    \label{fig:ExPUNITED}
\end{figure}

\begin{figure}[htb] 
    \centering
    \includegraphics[width=1.0\linewidth]{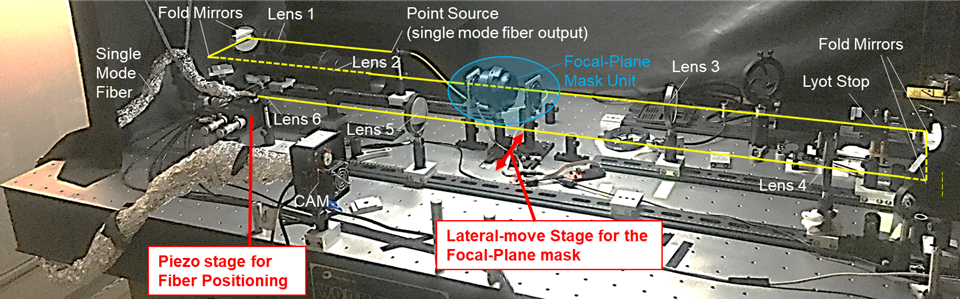}
    \caption{The picture of the real testbed of the experiment. The overlayed yellow line shows the approximated path of light in the setup. }
    \label{fig:FNP}
\end{figure}

\begin{figure}[htb] 
    \centering
    \includegraphics[width=0.7\linewidth]{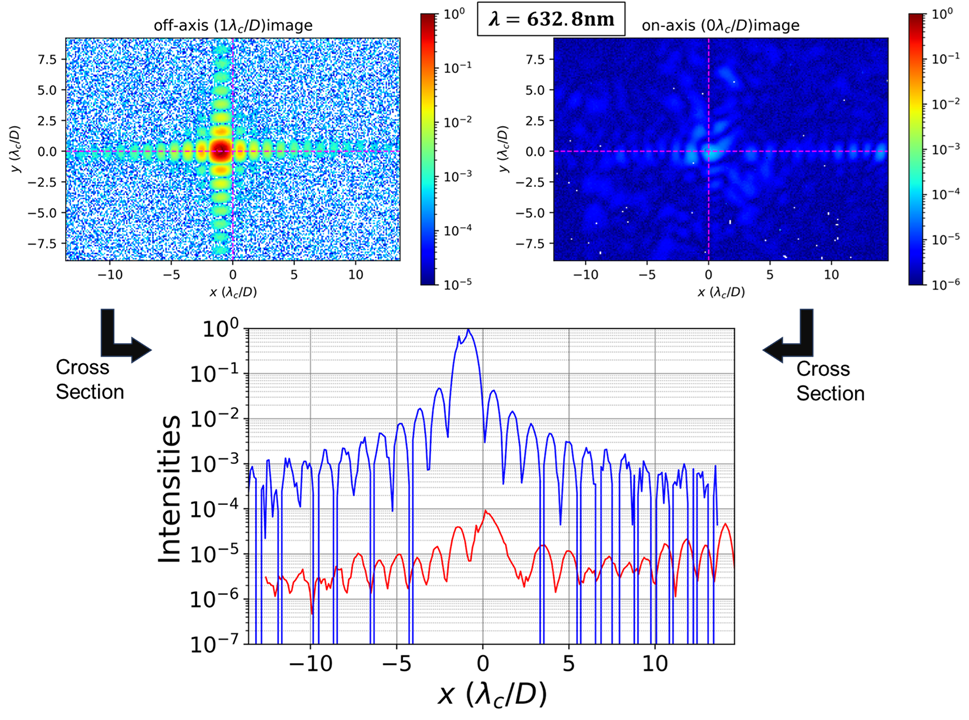}
    \caption{A typical coronagraphic output in the case of the center wavelength $\lambda_c$ and no Lyot
stop masks. The top panels exhibit the output image for off- (left) and on-axis (right) sources. The peak value of the off-axis image normalizes the image values. The bottom panel shows the cross-section across the horizontal magenta dashed lines in the top panels.}
    \label{fig:CR}
\end{figure}

\begin{figure}[htb] 
    \centering
    \includegraphics[width=0.7\linewidth]{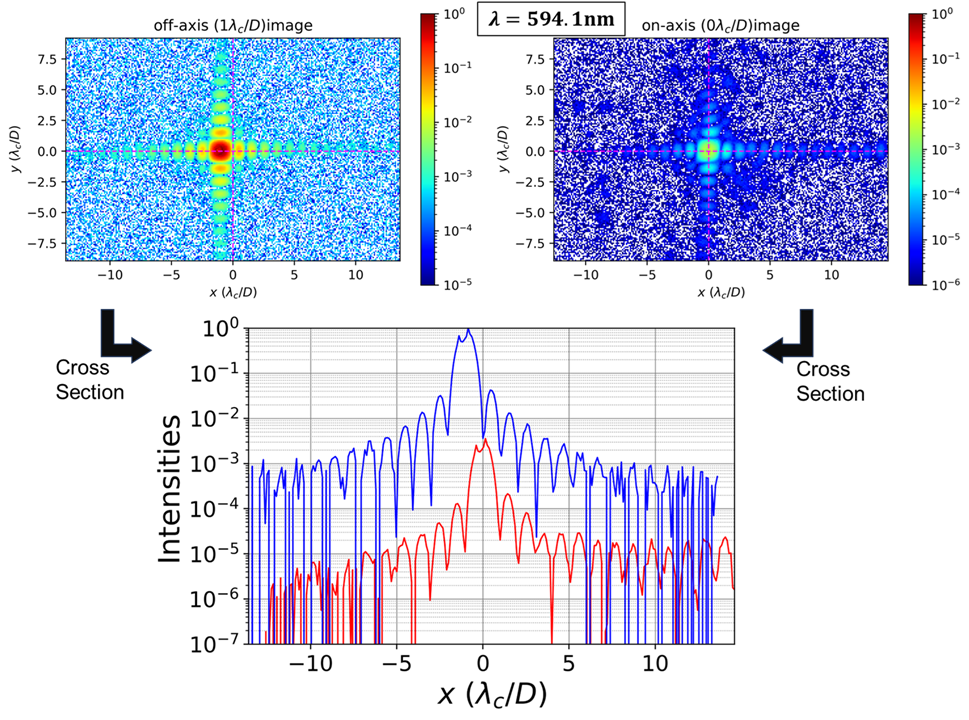}
    \caption{A typical coronagraphic output in the case of the wavelength $\lambda'$ that takes about 6-\% shorter than the center wavelength and no Lyot stop masks.}
    \label{fig:CY}
\end{figure}

Figure \ref{fig:ExPUNITED} illustrates the optical configuration of the present experiment.
Figure \ref{fig:FNP} shows a picture of the real testbed of the configuration illustrated in Figure \ref{fig:ExPUNITED}.
We use the experimental setup of the 1DDLC unit same as the previous study \citep[][]{2023PASP..135f4502I}, where we set the design center wavelength $\lambda_c$ of the focal-plane mask to 632.8 nm.
Figure \ref{fig:CR} indicates a typical coronagraphic output in the case of the center wavelength $\lambda_c$ and no Lyot stop masks. 
In the present experiment, we prepare two He-Ne gas lasers with different wavelengths ($\lambda_c=\mathrm{632.8\ nm}$ and $\lambda'=\mathrm{594.1\ nm}$) to test the wideband performance of the system.
 {Figure \ref{fig:CY} shows a typical coronagraphic output  in the case where $\lambda=\lambda'$. }
For the fiber nulling, we newly added a Lyot-plane mask, a relay optics ($\times1/100$), single-mode fiber actuated with a closed-loop 3-axis piezo stage, and a relay optics ($\times1$).

 {
To achromatically modulate the Lyot-plane phase according to an optimized mask pattern (Figure \ref{fig:SC}), the Lyot-plane mask consists of a linear polarizer (LP) and a half-wave plate (HWP) with a spatially patterned fast-axis direction to follow the following principle. 
The HWP folds the direction of the linear polarization of the coronagraph output concerning the orientation of the fast axes of the HWP. 
Hence, using the HWP with custom-patterned fast-axis orientation, we can achieve the arbitrary direction control of the linear polarization.
Depending on the direction of the input linear polarization, the linear polarizer after the HWP transmits only a single direction of the linear polarization with real-value amplitudes from minus one ($=e^{i \pi}$) to one ($=e^{0 \pi }$) such that the output spatial distribution realizes the desired mask pattern.
}
The focal-plane mask of the 1DDLC also uses the same implementation principle \citep[][]{2023PASP..135f4502I}.
In the present experiment, Photonic Lattice produces the custom pattern HWP.

Since the wavelengths of light determine the possible mode-field diameter of the single-mode fiber, we need to adjust the focal-plane magnification ratio of the optics to inject the light into the fiber effectively. 
Hence, we use a pair of achromatic lenses with nominal effective focal lengths of $\mathrm{400\ mm}$ (Edmund Optics, \#88595) and $\mathrm{4\ mm}$ (Thorlabs, SAC020-004-A) as a relay optics with a magnification ratio of 1/100. 
We set the design value of the spatial scale corresponding to the separation angles 1$\lambda_c/D$ on the fiber-injection focal plane to 2.5 $\mathrm{\mu m}$, where the symbol $D$ means the telescope aperture diameter.
Thus, the center core of the point spread function has the full width (from null point to null point) of 5.0 $\mathrm{\mu m}$ on this plane at the wavelength $\lambda_c$.
The used single-mode fiber (Thorlabs, 	
P1-S630Y-FC-2) has the nominal mode-field diameter of 4.2$\pm$0.5 $\mathrm{\mu m}$ for the wavelength of 630 nm that approximates the coronagraph's design-center wavelength $\lambda_c$ of 632.8 nm.   

In the present experiment, we actuate the single-mode fiber using a piezo stage (Thorlabs, MAX311D/M) with the closed-loop piezo controller (Thorlabs, BPC303).
Figure \ref{fig:ExPUNITED} shows a picture of the fiber-nulling unit. 
The present closed-loop control of the piezo actuator compensates for the hysteresis and the creep phenomenon. 
 {However, it leaves the positional variations due to the thermal shrinking of the piezo stage structure and/or surrounding optomechanics such as the relay optics.}
As reference data, we evaluated the positional instability of the piezo stage to about P-V 0.5 $\mathrm{\mu m}$ in 10 hours (Appendix \ref{piezo}). 

\subsection{Data Acquisition} \label{sec:meth2}
\subsubsection{On- and Off-axis Modes}
\begin{figure}[htb]  
    \centering
    \includegraphics[width=0.6\linewidth]{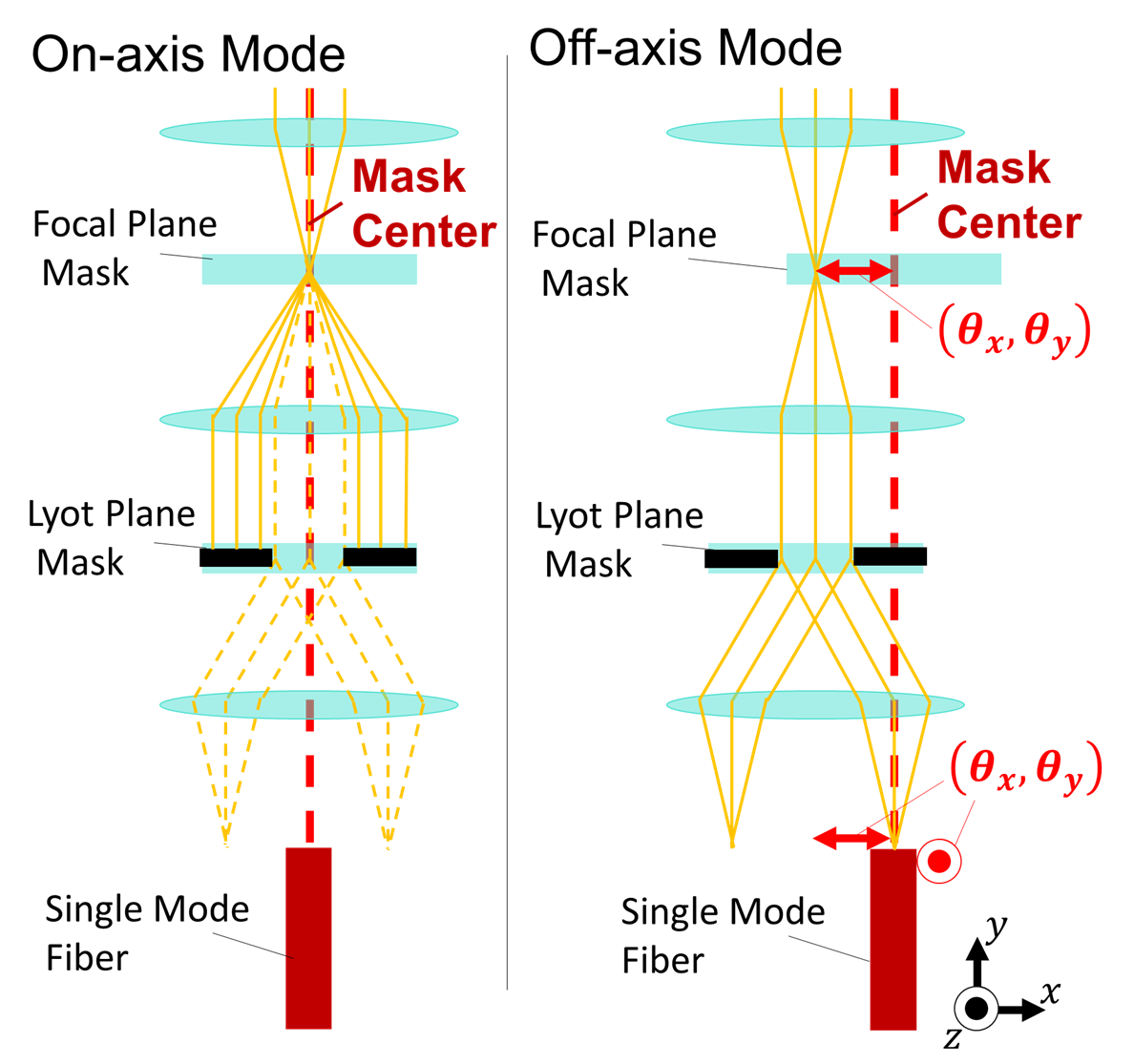}
    \caption{Definition of the data-acquisition modes. (left) On-axis mode. In the on-axis mode, we locate the center of each optical element on the optical axis. (right) Off-axis mode. In the off-axis mode, we laterally move the focal-plane mask and the single-mode fiber from the on-axis mode by the distances that correspond to the separation angles of $(\theta_x,\theta_y)$ to simulate the $(\theta_x,\theta_y)$-separated sources. In practice, we move the single-mode fiber parallel (x-axis) and perpendicular (z-axis) to the paper surface by $(\theta_x,\theta_y)$ because the Lyot-plane mask spit the beam along the direction not only parallel but also perpendicular to the paper surface. Note that the direction of the separation angle $\theta_y$ of the light source coincides with the direction of the z-axis of the piezo stage. In the experiment, we are not able to accurately designate these modes due to the thermal instability of the testbed, therefore we conduct fiber scans and select the data that approximates these modes later. The axes of 3-dimensional Cartesian coordinates at the right bottom indicate the driving axes of the piezo stage.}
    \label{fig:OnAndOff}
\end{figure}

\begin{figure}[htb] 
    \centering
    \includegraphics[width=0.7\linewidth]{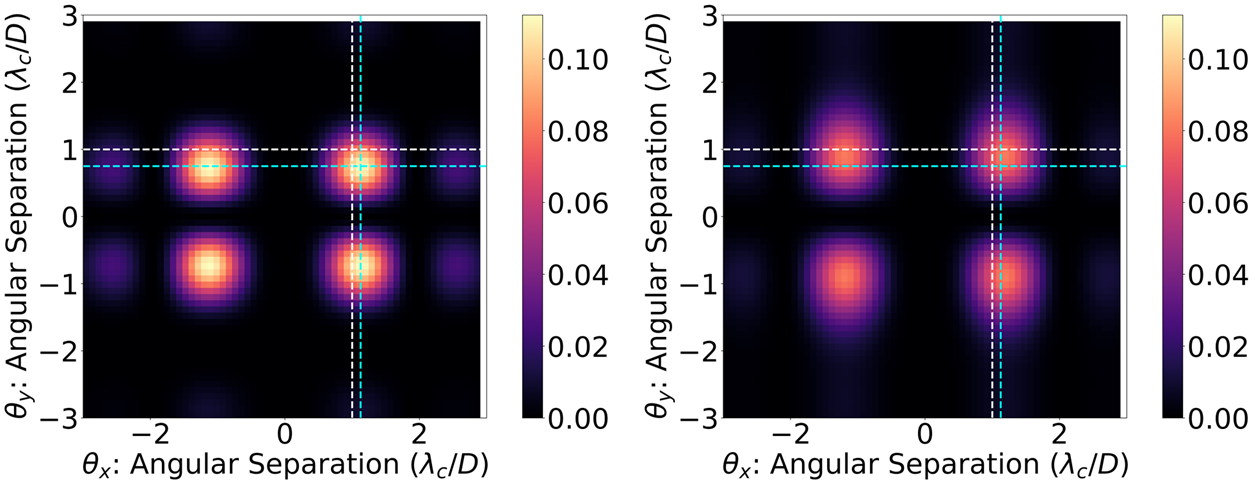}
    \caption{Simulation results (color-map value) for throughputs of the combined nuller in the case of the different two-dimensional positions of light sources (the horizontal and the vertical axes). The left panel indicates the case for an ideal single-mode fiber \citep[][]{2024AJ....167..235I}. The right panel shows the case for the nominal one used in the present experiment  {(off-the-shelf products that are not perfectly optimized for the experiment)}. The cross points of the white and cyan dashed lines express the  {positions} where $(\theta_x,\theta_y)=(1.00\lambda_c/D,1.00\lambda_c/D)$ and where $(\theta_x,\theta_y)=(1.13\lambda_c/D,0.75\lambda_c/D)$, respectively.}
    \label{fig:SIMTP}
\end{figure}

Here, we introduce two optical-setup modes (Figure \ref{fig:OnAndOff}) to define the measurements.
 {
In this experiment, we switch these modes by shifting the focal plane mask and single-mode fiber laterally, rather than the position of the light source, to avoid eccentric aberrations in the test bed lens. 
}
(i) In the on-axis mode, we locate the coronagraph's focal-plane mask and the single-mode fiber on the optical axis to test the on-axis source injection to  {the whole coronagraphic system that consists of the coronagraph and fiber-nulling units.} 
(ii) In the off-axis mode, we locate the coronagraph's focal-plane mask and the single-mode fiber on the position corresponding to the separation angles of $(\theta_x,\theta_y)$ to test the on-axis source injection to  {the whole coronagraphic system.}
In practice, we fix the values of $(\theta_x,\theta_y)$ to $(1.00\lambda_c/D,1.00\lambda_c/D)$
This value comes from the fact that the simulated throughput (Figure \ref{fig:SIMTP}) of the combined system has a peak around this separation.

\subsubsection{Fiber Scan}
 To place the fiber at the desired position even during thermal instability, we adopt a fiber scan with equal movements of the fiber position over an adjusted field of view size and select the approximately desired data later. 
 Due to the thermal instability, the absolute fiber positioning may include inaccuracies of about 0.5-$\mathrm{\mu m}$ (0.2$\lambda_c/D$) level when the fiber scanning takes a few hours (Appendix \ref{piezo}).
To produce a fiber output, we conduct photometry (over $20\times 20$ pixels) of image data after dark subtraction and defective pixel removal to the raw image data from the CCD camera (SBIG, ST-402ME).
In the off-axis mode (Figure \ref{fig:OnAndOff}), we conduct a simple scanning sequence with an exposure time of 40 ms for each scan point.

 In contrast, the high-resolution scan in the on-axis mode requires an optimized way of scanning because it needs long exposure times (e.g., 100 s) to produce the data with a sufficient signal-to-noise ratio. 
However, short scan time secures less systematic error from the thermal instability of the testbed structure (Appendix \ref{fpm}).
To address the trade-off between random and systematic errors, we need to allocate proper exposure times for all the individual scan points.
Hence, we adopt a deep-scan algorithm (Appendix \ref{korekara}) for the high-resolution scan in the on-axis mode. 
We executed the mid-resolution scan at the wavelength $\lambda=\lambda'=594.1\mathrm{nm}$ as a preparation for the fine-resolution scan to determine the fine-scan area.
The single mid- and fine-resolution scan requires about 20 and 200 min, respectively.
Since the fine-resolution scan takes a few hours, the absolute position of the fiber scanning image may include inaccuracies of about 0.5-$\mathrm{\mu m}$ level.

\subsection{Data Analysis}
 To evaluate the achieved contrast, we assume that the minimum values of the fine-resolution-scan image arise when the setup achieves the closest condition to the on-axis mode.
Thus, we have to estimate the minimum values in the fine-resolution-scan image with some noise.
The noise in the resultant fine-resolution deep scan image mainly contains dark-subtraction error, which comes from the temporal difference between dark- and light-exposure frames for the dark subtraction.
The dark-current difference between dark- and light-exposure frames derives from the temperature instability of the CCD with an air-cooled Peltier element, thus this noise should behave as a systematic noise rather than a random noise in principle.

\begin{figure}[htb] 
    \centering
    \includegraphics[width=0.5\linewidth]{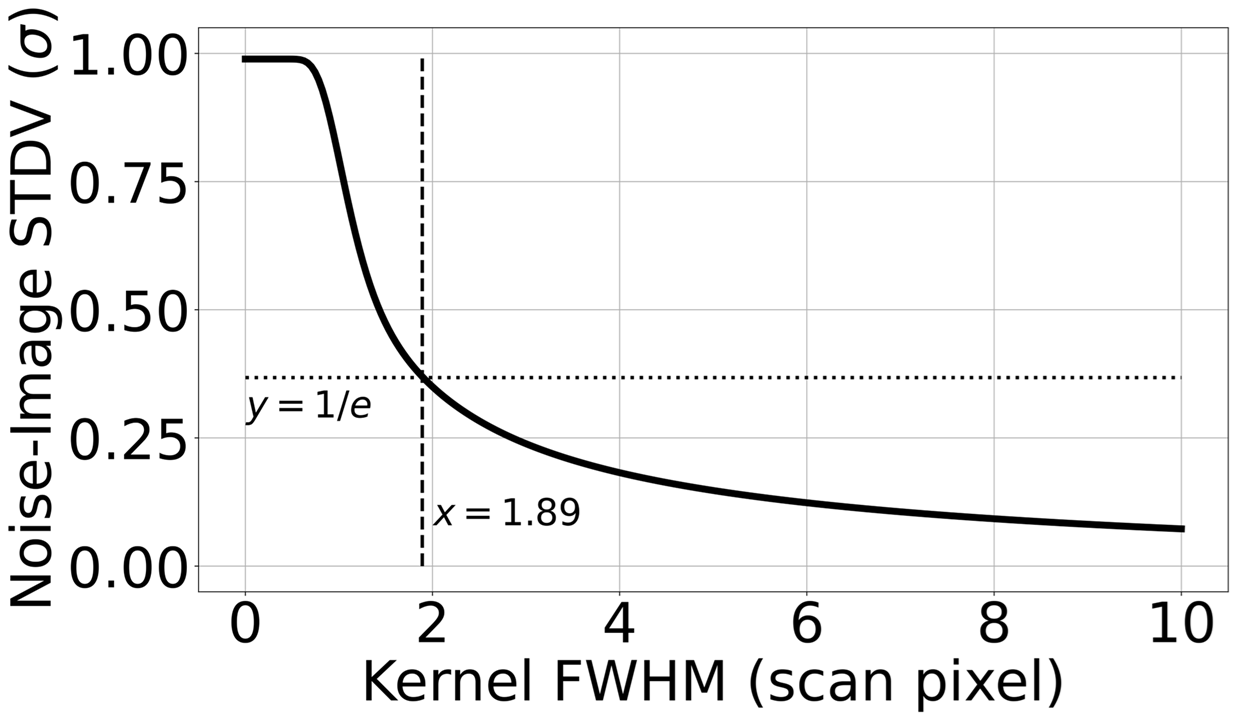}
    \caption{The simulation result for the relative noise standard deviation (STDV) of the Gaussian-filtered noise images (the vertical axis). The horizontal axis means the different kernel's full widths of half maximum (FWHM) utilized for the Gaussian filtering. }
    \label{fig:NG}
\end{figure}

 {Nevertheless, in the present measurement, the dark-subtraction error appears as a dominant random-noise factor in the resultant fiber scan images because all the individual point scans experience the well-randomized temporal temperature variation of the detector with an air-cooled Peltier element.
The resultant dark-subtraction error approximates white noise on the scan image. 
Hence, we need to operate the Gaussian filtering with an optimized kernel width to the scanned image before the minimum-value determination.
To optimize the kernel width, we simulate the noise reduction for the assumed white noise image array with the same array size as the one ($20\times20$ pixels) in the present scan image.
Figure \ref{fig:NG} shows the simulation result of the relative noise reduction rate for the different kernel widths of the Gaussian filtering.
Considering this simulation result, we adopt a kernel's full width of half maximum (FWHM) of 1.89 scan pixels, with which the standard deviation of the noise image takes $1/e$ times the one of the original noise image, as the optimized value.  }

\section{Result} \label{sec:resu}
\begin{figure}[htb] 
    \centering
    \includegraphics[width=0.7\linewidth]{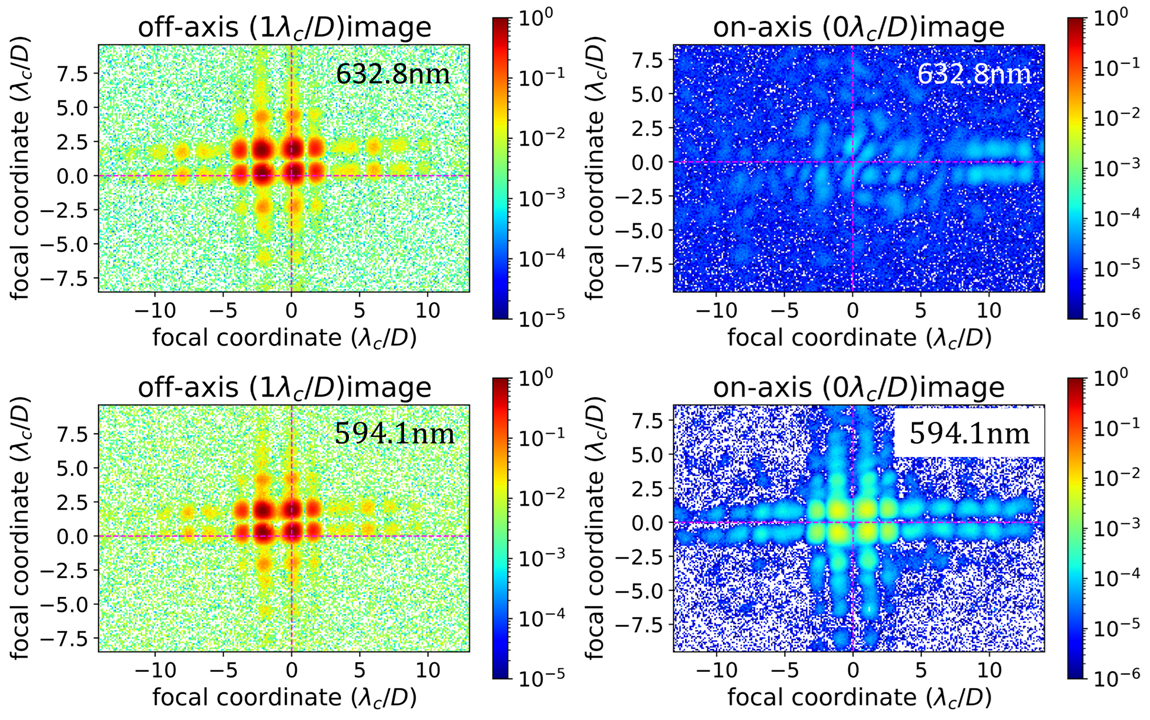}
    \caption{The output focal images from the coronagraph with the Lyot-stop mask just before the fiber scanning of Figure \ref{fig:YO}. The left and right panels correspond to the cases with off- and on-axis modes, respectively. The top and bottom rows show the cases for different wavelengths of the light sources. The colormaps indicate the intensity normalized by the peak of the off-axis image. The vertical magenta dashed lines show the center of the focal plane mask. The cross point of the vertical and horizontal magenta dashed lines denotes the place where we should locate the single-mode fiber for each mode.  }
    \label{fig:1211}
\end{figure}

Here, we observe the result of the raw contrast mitigation confirmed by the present study to see whether the spectral bandwidth expanded thanks to the serially connecting the fiber nulling after the 1DDLC.
Figure \ref{fig:1211} shows the coronagraphic focal outputs of the 1DDLC with the Lyot-stop mask as a reference. 
These images come as a result of the dark subtraction from the raw CCD data.
\begin{figure}[htb] 
    \centering
    \includegraphics[width=1.0\linewidth]{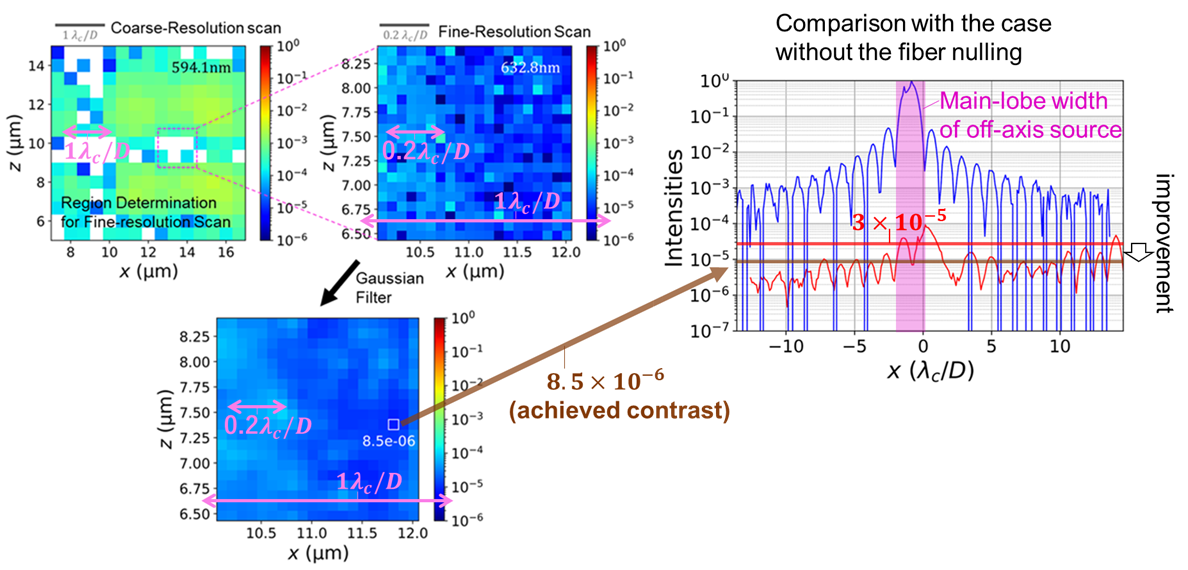}
    \caption{The left three panels show the fiber scan images showing the measured contrast for the wavelength of 632.8 nm (the coronagraph center wavelength) in the neighborhood of the on-axis mode (Figure \ref{fig:OnAndOff}). The horizontal and the vertical axes denote command values to the piezo actuator with a closed loop control (the x- and z-axes in Figure \ref{fig:ExPUNITED} and \ref{fig:OnAndOff}, respectively). The color maps mean the fiber outputs normalized by the fiber-output intensity in the off-axis mode (Figure \ref{fig:OnAndOff}). The top left panel shows the coarse resolution scan (using a yellow laser) just before conducting the fine resolution scan. The magenta dashed square indicates the area for the fine-resolution scan indicated in the top right one among the left three panels. The white pixels mean that the fiber-output data have values smaller than zero due to measurement error. The bottom panel shows the Gaussian-filtered images of the top right one among the left three panels. The white cube shows the location of the minimum value over the filtered scan image. The right panel compares the achieved contrast indicated in the left panels with the case without the fiber nulling indicated in Figure \ref{fig:CR}. The magenta-filled region expresses the main-lobe width of the off-axis source, over which we integrate the light intensity using reasonable aperture photometry. The red horizontal line shows the value of the normalized intensity of the on-axis source (red curve) averaged over the magenta-filled region. Thus, the red horizontal line indicates the achieved contrast mitigation ability in the case without the fiber nulling. The brown horizontal line shows the achieved contrast of the system consisting of the 1DDLC and the fiber nulling in the present study.  }
    \label{fig:RO}
\end{figure}
\begin{figure}[htb] 
    \centering
    \includegraphics[width=1.0\linewidth]{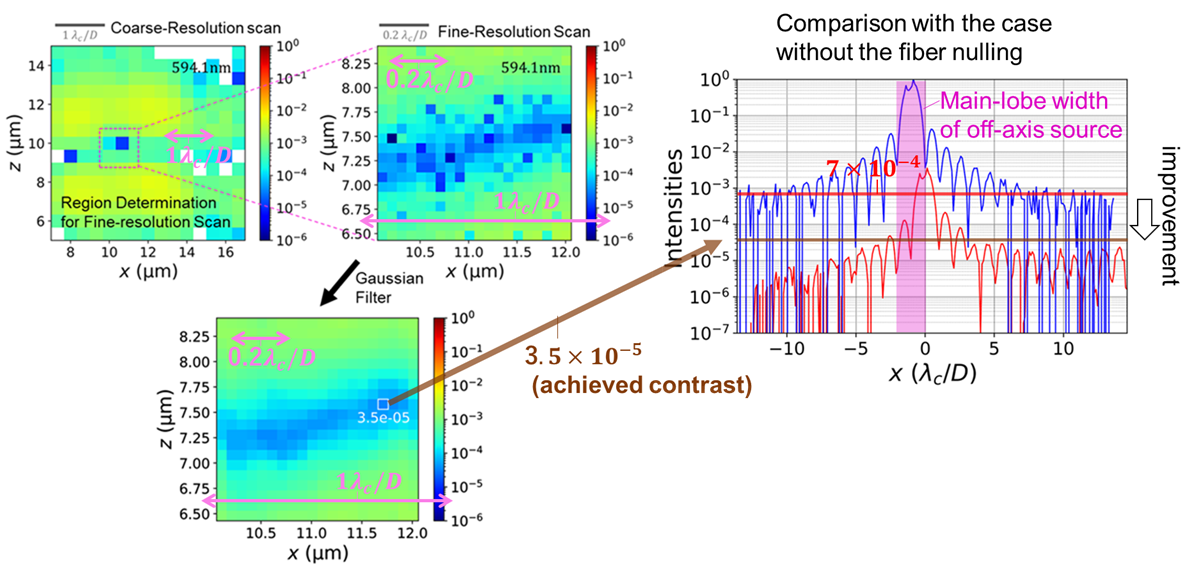}
    \caption{The left three panels show the fiber scan images showing the measured contrast for the wavelength of 594.1 nm (about 6\% shorter than the coronagraph center wavelength) in the neighborhood of the on-axis mode (Figure \ref{fig:OnAndOff}).  The horizontal and the vertical axes denote command values to the piezo actuator with a closed loop control (the x- and z-axes in Figure \ref{fig:ExPUNITED} and \ref{fig:OnAndOff}, respectively). The color maps mean the fiber outputs normalized by the fiber-output intensity in the off-axis mode (Figure \ref{fig:OnAndOff}). The top left panel shows the coarse resolution scan (using a yellow laser) just before conducting the fine resolution scan. The magenta dashed square indicates the area for the fine-resolution scan indicated in the top right  one among the left three panels. The white pixels mean that the fiber-output data have values smaller than zero due to measurement error. The bottom panel shows the Gaussian-filtered images of the top right panel. The white cube shows the location of the minimum value over the filtered scan image. The right panel compares the achieved contrast indicated in the left panels with the case without the fiber nulling indicated in Figure \ref{fig:CY}. The magenta-filled region expresses the main-lobe width of the off-axis source, over which we integrate the light intensity using reasonable aperture photometry. The red horizontal line shows the value of the normalized intensity of the on-axis source (red curve) averaged over the magenta-filled region. Thus, the red horizontal line indicates the achieved contrast mitigation ability in the case without the fiber nulling. The brown horizontal line shows the achieved contrast of the system consisting of the 1DDLC and the fiber nulling in the present study. }
    \label{fig:YO}
\end{figure}
 {Figures \ref{fig:RO} and \ref{fig:YO} show the results of the fiber scan near the on-axis mode normalized by the peak value of the scan in the off-axis mode for the case where $\lambda=\lambda_c=632.8\ \mathrm{nm}$ and where $\lambda=\lambda'=594.1\ \mathrm{nm}$), respectively. 
The top left panels of Figures \ref{fig:RO} and \ref{fig:YO} indicate mid-resolution scan images of the wavelength of $\lambda'$, which guide the region determination of the fine-resolution scans (top right panels).
The bottom panels of Figures \ref{fig:RO} and \ref{fig:YO} exhibit Gaussian-filtered fine-scan images.}

The minimum values of the bottom panels of Figures \ref{fig:RO} and \ref{fig:YO} correspond to the contrast-mitigation ability confirmed in the present study for $\lambda=\lambda_c=632.8\ \mathrm{nm}$ and $\lambda=\lambda'=594.1\ \mathrm{nm}$, respectively.
For the coronagraph's design center wavelength $\lambda_c$, the confirmed contrast-mitigation performance reaches $8.5\times10^{-6}$.
The experiment intended no contrast improvement at the center wavelength $\lambda_c$, but adding the fiber nuller shows a slight reduction from the value about $1\times 10^{-5}$ of the contrast-mitigation performance for the case with only the 1DDLC demonstrated in the previous study \citet[][]{2023PASP..135f4502I}.

For the wavelength $\lambda'$ that is about 6-\% shorter than $\lambda_{c}$, we confirmed the contrast mitigation ability of $3.5\times10^{-5}$.
Compared with Figure \ref{fig:CY} (the case with only the 1DDLC)   where the contrast mitigation performance is about $7\times 10^{-4}$, connecting the fiber nuller including the Lyot-stop mask reduced the value of the contrast-mitigation performance to about 1/20 times the value before the connecting. 
The resultant value of $3.5\times10^{-5}$ has an amount approximately equal to the value $3\times10^{-5}$ for the case with only the 1DDLC at the center wavelength $\lambda_c$. 
In addition, we can suppose that similar results will occur for the intermediate wavelengths between the tested wavelengths $\lambda_c$ and $\lambda'$ because the experimental setup should include no elements with a steep dependency for the wavelength of light.
Thus, we can say that the demonstrated spectral bandwidth reaches at least 6\% on one side remote from the design center wavelength (for the contrast-mitigation level of about $10^{-5}$). 

\section{Conclusion} \label{sec:conc}
 We conducted an experimental verification of a serially conjoined nuller that consists of a type (1DDLC) of Lyot coronagraph and a fiber nulling.
To place the fiber at the desired position even during thermal instability, we had to deploy the fiber scan to achieve precise fiber positioning through the post-selection of the data.
Because the dark-subtraction noise dominates the fine-resolution scan image, we had to operate the Gaussian filtering with an optimized kernel width to the fine-scan image before we determined which scan point corresponds to the best approximation to the desired fiber position.
Although the achieved contrast-mitigation level currently stays around $10^{-5}$, we suppose that a similar broadening of spectral bandwidths serves for the contrast-mitigation level lower than $10^{-5}$ in principle.

 We achieved the raw contrast of about $3.5\times10^{-5}$ for the wavelength 6-\% less than the design-center wavelength, suggesting that the combined system works robustly against the wide spectral bandwidth.
Although the current contrast-mitigation level currently stays around $10^{-5}$, a similar broadening of spectral bandwidths serves for the contrast-mitigation level lower than $10^{-5}$ in principle.
Future work needs to address the demonstration of this anticipation. 



\bibliography{example}{}
\bibliographystyle{aasjournal}




\appendix

\section{Stability of the piezo actuator stage with a closed-loop control}\label{piezo}
\begin{figure}[htb]
    \centering
    \includegraphics[width=0.5\linewidth]{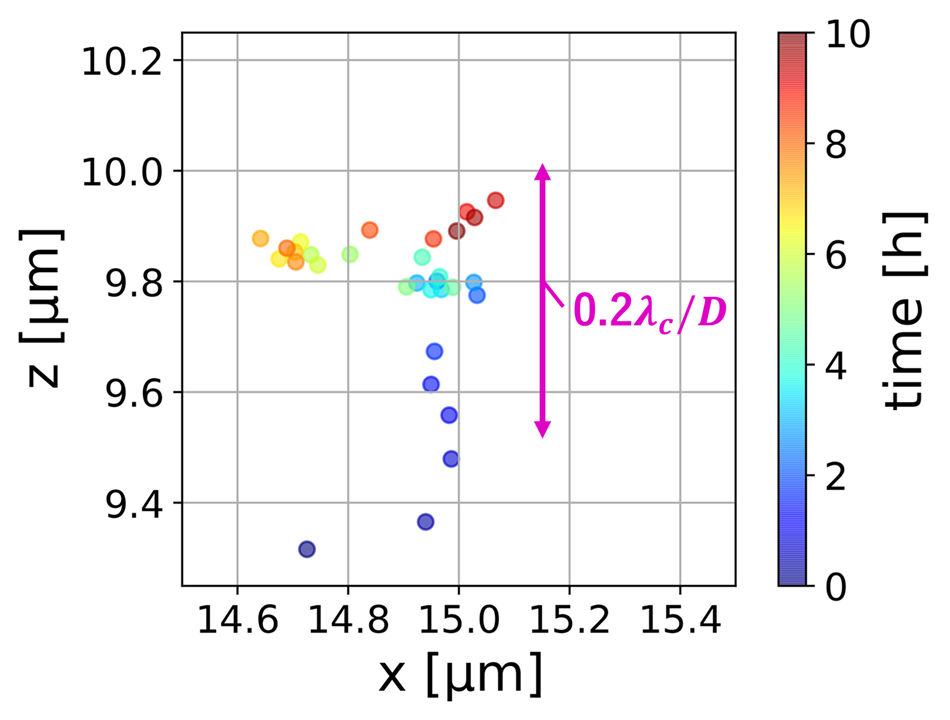}
    \caption{Example data concerning the stability of the closed-loop control piezo actuator stage {, which may include instability of the relay optics with the magnification ratio of 1/100.} Each plot means the center coordinates of the peaks in the fiber scan images under the condition that we intend to fix the incident beam.  {The data may include the incident-beam variation due to the instability of the relay optics with the magnification ratio of 1/100. }The different colors of markers denote the different elapsed time from the measurement start.}
    \label{fig:PS}
\end{figure}
Figure \ref{fig:PS} shows measurement data for the stability of the piezo-actuated stage with a closed-loop control. 
To obtain the data, we conduct fiber scans successively through 10 hours in the neighborhood of the off-axis mode (Figure \ref{fig:OnAndOff}) without sending any other commands to the stage.
The data points plotted in Figure \ref{fig:PS} indicate the intensity centers of the fiber-scan images.

\section{stability of the focal plane mask adjustment optomechanics}\label{fpm}
\begin{figure}[htb] 
    \centering
    \includegraphics[width=0.5\linewidth]{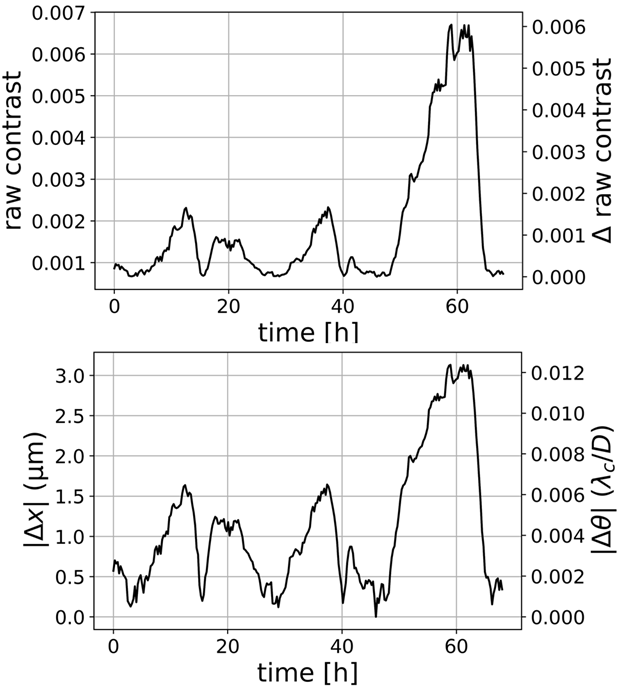}
    \caption{Example data for the stability of the focal plane mask adjustment optomechanics. The horizontal axes indicate the elapsed time after the measurement starts. (top) The left vertical axis indicates the raw contrast of the 1DDLC measured as the peak of the leak on the coronagraph output focal plane at different times. The right vertical axis offsets the left axis such that the minimum value of the data sequence for the raw contrast takes zero. (bottom) The left vertical axis shows the amount of lateral movement of the focal-plane mask estimated through the assumption that all the increases in the raw contrast from its minimum value come from the lateral shift of the focal-plane mask. The right vertical axis translates the left axis to the amount of the tilt aberration due to the mismatch of the mask center and the source position. }
    \label{fig:TCDX}
\end{figure}
The top panel of Figure \ref{fig:TCDX} shows long-term stability in the raw contrast of the 1DDLC setup. 
Measured profiles of the focal leak suggest that the instability mainly comes from the thermal instability of the adjustment opt-mechanics that consists of a rail carrier positioner for dovetail rails (RCN and RCN1, Thorlabs) and actuates the focal plane mask.
The bottom panel of Figure \ref{fig:TCDX} assumes that all the increases in the raw contrast from its minimum value come from the lateral shift of the focal-plane mask (tilt aberration).
\section{Deep-scan algorithm for the high-resolution scan}\label{korekara}
Each point scan consists of the following sequences:
\begin{enumerate}
    \item Set the iteration counter $n$ to 0.
    \item Obtain the fiber output $I(n)$ with a exposure time $\tau_{\mathrm{exp}}$, where
 {$\tau_{\mathrm{exp}}=(40\ \mathrm{ms})e^{\log(2)n}$. }
    \item If the outputs $I(n-1)$ and $I(n)$ exist and take non-negative values, complete the point scan, convert the output value to the per-unit-time amount, and go to the next point scan. Otherwise, change the iteration counter $n$ to $n+1$ and repeat the sequence from (ii).
\end{enumerate}
Thanks to this algorithm, we can take much time for the scan points with small powers of outputs, resulting in relatively homogeneous signals-to-noise ratios over the scan field and less waste of measurement time.

\end{document}